\documentclass[aps,pr,superscriptaddress,preprintnumbers,nofootinbib,10pt]{revtex4-2}

\usepackage{multirow}
\usepackage{amsmath}
\usepackage{amssymb}
\usepackage[dvipdf,dvips]{graphicx}
\usepackage{color}
\usepackage{hyperref}
\usepackage{url}
\usepackage{slashed}
\usepackage{subfigure}
\usepackage[usenames,dvipsnames]{xcolor}
\usepackage{amsmath}
\usepackage{amsfonts}
\usepackage{float} 
\usepackage{amssymb}
\usepackage{epsfig}
\usepackage{graphics}
\usepackage{euscript}
\usepackage{slashed}
\usepackage{epstopdf}
\usepackage[utf8]{inputenc}
\allowdisplaybreaks
\usepackage[normalem]{ulem}
\usepackage{pifont}
\usepackage{dsfont}
\usepackage{MnSymbol}
\usepackage{verbatim}
\usepackage{graphicx}
\usepackage{latexsym}
\usepackage{tikz-feynman}
\usepackage{bbold}

\begin{document}

	\title{{\bf Gluing together Quantum Field Theory and Quantum Mechanics: a look at the Bell-CHSH inequality} }

	
	
	\author{M. S.  Guimaraes}\email{msguimaraes@uerj.br} \affiliation{UERJ $–$ Universidade do Estado do Rio de Janeiro,	Instituto de Física $–$ Departamento de Física Teórica $–$ Rua São Francisco Xavier 524, 20550-013, Maracanã, Rio de Janeiro, Brazil}
	
	\author{I. Roditi} \email{roditi@cbpf.br} \affiliation{CBPF $-$ Centro Brasileiro de Pesquisas Físicas, Rua Dr. Xavier Sigaud 150, 22290-180, Rio de Janeiro, Brazil}
	
	\author{S. P. Sorella} \email{silvio.sorella@gmail.com} \affiliation{UERJ $–$ Universidade do Estado do Rio de Janeiro,	Instituto de Física $–$ Departamento de Física Teórica $–$ Rua São Francisco Xavier 524, 20550-013, Maracanã, Rio de Janeiro, Brazil}

	\begin{abstract}

The Bell-CHSH inequality in the vacuum state of a relativistic scalar quantum field is revisited by making use of the Hilbert space ${\cal H} \otimes {\cal H}_{AB}$, where ${\cal H}$ and ${\cal H}_{AB}$ stand, respectively, for the Hilbert space of the scalar field and of a generic bipartite quantum mechanical system. The construction of Hermitian, field-dependent, dichotomic operators is devised as well as the Bell-CHSH inequality. Working out the $AB$ part of the inequality, the resulting Bell-CHSH correlation function for the quantum field naturally emerges  from unitary Weyl operators. Furthermore, introducing a Jaynes-Cummings type Hamiltonian accounting for the interaction between the scalar field and a pair of qubits, the quantum corrections to the Bell-CHSH inequality in the vacuum state of the scalar field are evaluated till the second order in perturbation theory. 

	\end{abstract}

	\maketitle

	\section{introduction}\label{intro}
	
	Since the fundamental work by \cite{SW1,Summers87b,Summers:1987ze}, the study of the Bell-CHSH inequality \cite{Bell64,CHSH69} in relativistic Quantum Field Theory has witnessed an increasing interest, both at theoretical  and phenomenological level   \cite{DeFabritiis:2023tkh,Barr:2024djo}. \\\\In the present work, borrowing ideas from \cite{Summers:1987ze} and from the investigation of the Unruh-De Witt detectors \cite{Reznik:2002fz,Tjoa:2022vnq,Tjoa:2022lel,Lima:2023pyt,Barbado:2020snx,Foo:2020xqn,Verdon-Akzam:2015tma,Guedes:2024tcq}, we perform an analysis of the Bell-CHSH inequality by looking at a system whose Hilbert space is given by ${\cal H} \otimes {\cal H}_{AB}$, where ${\cal H}$ is the Hilbert space of a relativistic scalar quantum field $\varphi(t,x)$ and ${\cal H}_{AB}$ denotes the Hilbert space of a bipartite quantum mechanical system. The space ${\cal H}_{AB}$ can be either of finite or infinite dimension. That construction is very close to the formulation of the Jaynes-Cummings model, see \cite{Larson:2022jvs}, describing features of the interaction between electromagnetic fields and atoms. \\\\There are several advantages in looking at the Hilbert space  ${\cal H} \otimes {\cal H}_{AB}$, namely:
	\begin{itemize}
	
	\item the construction of Hermitian, field-dependent, dichotomic operators can be outlined simply and elegantly. 
	
	\item the Bell-CHSH inequality in the vacuum state of the scalar field $\varphi(t,x)$ can be accessed by working out the quantum mechanical part of the whole inequality formulated on ${\cal H} \otimes {\cal H}_{AB}$.
	
	\item the unitary Weyl operators 
	\begin{equation} 
	W_f = e^{i \varphi(f)} \;, \label{wop}
	\end{equation}
	where $\varphi(f)$ is the smeared field \cite{Haag92}, emerge in a natural way in the expression for the Bell-CHSH correlator for the quantum field. 
	
	\item A Jaynes-Cummings type Hamiltonian describing the interaction between $\varphi(t,x)$ and a pair of qubits can be written down and employed to investigate the existence of quantum corrections to the Bell-CHSH inequality for the quantum field. An explicit calculation till the second order in perturbation theory will be reported. 
	
	\end{itemize} 
	
	\noindent In the following, we shall elaborate in detail on all above-mentioned issues

\section{Field quantization, Tomita-Takesaki modular theory  theory and the von Neumann algebra of the Weyl Operators}\label{TTW}

As the work is meant to be self-contained, it is worth starting by reminding a few basic notions of  Quantum Field Theory and of the Von Neumann algebra built out from the unitary Weyl operators\footnote{See ref.\cite{DeFabritiis:2023tkh} for a more detailed account.}. We shall follow \cite{DeFabritiis:2023tkh} and consider a real free scalar field in four-dimensional Minkowski space-time:
\begin{eqnarray} 
\varphi(t,x) & = & \int \frac{d^3p}{(2\pi)^3} \frac{1}{2 \omega_p} \left( e^{-ipx} a_p + e^{ipx} a^{\dagger}_p \right) \;, \qquad p_0=\omega_p =\sqrt{{\vec p}^2 + m^2} \;, \nonumber  \\
\left[ a_p, a^{\dagger}_q \right]  & = &  {(2\pi)^3} 2\omega_p \; \delta^3(p-q) \;, \qquad [a_p, a_q] = [a^{\dagger}_p, a^{\dagger}_q] = 0 \;. \label{qf}
\end{eqnarray}
For the  expression of the causal Pauli-Jordan distribution $\Delta_{PJ}(x-y)$ one gets
\begin{eqnarray}
[\varphi(x), \varphi(y)] & = & i \Delta_{PJ}(x-y) \nonumber \\
    i \Delta_{PJ}(x-y) \! &=&\!\! \int \!\! \frac{d^4k}{(2\pi)^3} \varepsilon(k^0) \delta(k^2-m^2) e^{-ik(x-y)} \;, \nonumber \\
    \label{PJ}
\end{eqnarray}
with $\varepsilon(x) \equiv \theta(x) - \theta(-x)$.\\\\$\Delta_{PJ}(x-y)$ is Lorentz invariant and vanishes when $x$ and $y$ are space-like
\begin{equation} 
\Delta_{PJ}(x-y) = 0 \;, \quad {\rm for} \quad (x-y)^2<0 \;. \label{spl}
\end{equation}
Let $\mathcal{O}$ be an open region of the Minkowski spacetime and let $\mathcal{M}(\mathcal{O})$ be the space of smooth test functions $\in \mathcal{C}_{0}^{\infty}(\mathbb{R}^4)$ with support contained in $\mathcal{O}$:
\begin{align} 
	\mathcal{M}(\mathcal{O}) = \{ f \, \vert supp(f) \subseteq \mathcal{O} \}. \label{MO}
\end{align}
One introduces the symplectic complement \cite{SW1,Summers87b} of $\mathcal{M}(\mathcal{O})$ as 
\begin{align} 
	\mathcal{M'}(\mathcal{O}) = \{ g \, \vert  \Delta_{PJ}(g,f) = 0, \; \forall f \in \mathcal{M}(\mathcal{O}) \}, \label{MpO}
\end{align}
that is, $\mathcal{M}(\mathcal{O})$ is given by the set of all test functions for which the smeared Pauli-Jordan expression $\Delta_{PJ}(f,g)$  vanishes for any $f$ belonging to $\mathcal{M}(\mathcal{O})$ 
\begin{equation} 
\left[ \varphi(f), \varphi(g) \right] =  i  \Delta_{PJ}(f,g) \;. \label{smpj}
\end{equation} 
The symplectic complement $\mathcal{M'}(\mathcal{O})$ allows one to recast causality, eq.\eqref{spl}, as \cite{SW1,Summers87b}
\begin{align}
    \left[ \varphi(f), \varphi(g) \right] = 0,
\end{align}
whenever $f \in \mathcal {M}(\mathcal{O})$ and $g \in \mathcal {M'}(\mathcal{O})$. \\\\We proceed by introducing the unitary Weyl operators \cite{SW1,Summers87b,DeFabritiis:2023tkh},  obtained by exponentiating the smeared field 
\begin{equation} 
W_{h} = e^{i {\varphi}(h) }. 
\label{Weyl}
\end{equation}
These operators give rise to the so-called Weyl algebra
\begin{align} \label{algebra} 
	W_{f} W_{g}   &= e^{ - \frac{i}{2} \Delta_{\textrm{PJ}}(f, g)}\; W_{(f+g)}, \nonumber \\
	W_{f}^{\dagger} W_{f} &= 	W_{f} W_{f}^{\dagger} = 1, \nonumber \\ 
	W^{\dagger}_{f} &=  W_{(-f)}.	 
\end{align} 
For $f$ and $g$ space-like, the Weyl operators $W_{f}$ and $W_{g}$ commute. The vacuum expectation value of $W_h$ is found 
\begin{equation} 
\langle 0 \vert  W_{h}  \vert 0 \rangle = \langle 0 \vert  W_{(-h)}  \vert 0 \rangle = \; e^{-\frac{1}{2} {\lVert h\rVert}^2}, 
\label{valueW}
\end{equation} 
where $\vert\vert h \vert\vert^2 =  \langle h \vert h \rangle$ and 
\begin{equation} 
\langle f \vert g \rangle  =  \int \frac{d^3k}{(2\pi)^3} \frac{1}{2\omega_k} f(\omega_k,\vec{k})^{*} g(\omega_k,\vec{k}) \;, \qquad
f(k_0, \vec{k}) = \int d^4x \; e^{ikx} f(x) \;,  \label{inner}
\end{equation} 
is the Lorentz invariant inner product between the test functions $(f,g)$\footnote{For $\omega_k$ we have the usual relation $\omega^2_k = k^2+m^2$. } \cite{SW1,Summers87b,DeFabritiis:2023tkh}. Taking products and linear combinations of the Weyl operators defined on $\mathcal{M}(\mathcal{O})$, one gets a von Neumann algebra $\mathcal{A}(\mathcal{M})$. Moreover, from the Reeh-Schlieder theorem \cite{Haag92, Witten18}, it turns out that the vacuum state $\vert 0 \rangle$ is cyclic and separating for the von Neumann algebra $\mathcal{A}(\mathcal{M})$. As such, one can make use of the Tomita-Takesaki modular theory \cite{Bratteli97,Witten18,SW1,Summers87b,DeFabritiis:2023tkh} and introduce the anti-linear unbounded operator $S$ whose action on  $\mathcal{A}(\mathcal{M})$ is defined as
\begin{align} 
	S \; a \vert 0 \rangle = a^{\dagger} \vert 0 \rangle, \qquad \forall a \in \mathcal{A}(\mathcal{M}),
    \label{TT1}
\end{align}  
from which it follows that $S^2 = 1$ and $S \vert 0 \rangle = \vert 0 \rangle$. Employing the polar decomposition \cite{Bratteli97,Witten18,SW1,Summers87b,DeFabritiis:2023tkh}, one gets
\begin{align}
    S = J  \Delta^{1/2},
    \label{PD}    
\end{align} 
where $J$ is anti-unitary  and $\Delta$ is positive and self-adjoint. These  operators satisfy the following properties \cite{Bratteli97,Witten18,SW1,Summers87b,DeFabritiis:2023tkh}:
\begin{align} 
	J \Delta^{1/2} J &= \Delta^{-1/2}, \quad \,\,	\Delta^\dagger = \Delta, \nonumber \\
	S^{\dagger} &= J \Delta^{-1/2},  \,\,\,\,\, J^{\dagger} = J, \nonumber \\
	\Delta &= S^{\dagger} S,  \quad \,\,\,\,\,\, J^2 = 1.
    \label{TTP}
\end{align}
According to the Tomita-Takesaki theorem \cite{Bratteli97,Witten18,SW1,Summers87b,DeFabritiis:2023tkh}, one has that   $J \mathcal{A}(\mathcal{M}) J = \mathcal{A}'(\mathcal{M})$, that is, upon conjugation by the operator $J$, the algebra $\mathcal{A}(\mathcal{M})$ is mapped into its commutant $\mathcal{A'}(\mathcal{M})$, namely: 
\begin{equation} 
\mathcal{A'}(\mathcal{M}) = \{ \; a' \, \vert \; [a,a']=0, \forall a \in \mathcal{A}(\mathcal{M}) \;\}.
    \label{commA}
\end{equation} 
The Tomita-Takesaki modular theory is particularly suited for analyzing the Bell-CHSH inequality within relativistic Quantum Field Theory \cite{SW1,Summers87b}. As discussed in  \cite{DeFabritiis:2023tkh},  it gives a way of constructing in a purely algebraic way Bob's operators from Alice's ones by using the modular conjugation $J$. That is, given Alice's operator $A_f$, one can assign the operator $B_f = J A_f J$ to Bob, with the guarantee that they commute since, by the Tomita-Takesaki theorem, the operator $B_f = J A_f J$ belongs to the commutant $\mathcal{A'}(\mathcal{M})$ \cite{DeFabritiis:2023tkh}. \\\\Following  \cite{Rieffel77,Eckmann73},  the Tomita-Takesaki modular theory can be lifted to the space of test functions. One notices that, when equipped with the Lorentz-invariant inner product $\langle f \vert g\rangle$, eq.\eqref{inner}, the set of test functions gives rise to a complex Hilbert space $\mathcal{F}$ which enjoys several features. More precisely,  it turns out that the subspace $\mathcal{M}$ is a standard   subspace for $\mathcal{F}$ \cite{Rieffel77}, meaning that:  i) $\mathcal{M} \cap i \mathcal{M} = \{ 0 \}$; ii) $\mathcal{M} + i \mathcal{M}$ is dense in $\mathcal{F}$. From \cite{Rieffel77}, for such subspaces it is possible to set a modular theory analogous to that of the Tomita-Takesaki. One introduces an operator $s$ acting on $\mathcal{M} + i\mathcal{M}$ as
\begin{align}
    s (f+ih) = f-ih. \;, 
    \label{saction}
\end{align}
for $f,h \in \mathcal{M}$. Notice that $s^2 = 1$. Using the  polar decomposition, one has:  
\begin{align}
    s = j \delta^{1/2},  \label{jd}
\end{align}
where $j$ is an anti-unitary operator and $\delta$ is  positive and self-adjoint.  Similarly to the operators $(J,\, \Delta)$, the  operators $(j,\, \delta)$ fulfill  the following properties \cite{Rieffel77}:
\begin{align}
    j \delta^{1/2} j &= \delta^{-1/2}, \,\,\,\,\,\,  \delta^\dagger = \delta\nonumber \\
    s^\dagger &= j \delta^{-1/2}, \,\,\, j^\dagger = j \nonumber \\
    \delta &= s^\dagger s, \,\,\,\,\,\,\,\,\,\,\, j^2=1.
\end{align}
Moreover,  a test function $f$ belongs to $\mathcal{M}$ if and only if \cite{Rieffel77}
\begin{equation} 
s f = f \;. 
    \label{sff}
\end{equation}
Similarly,  one has that $f' \in \mathcal{M}'$ if and only if $s^{\dagger} f'= f'$. \\\\The lifting of the action of the operators $(J, \Delta)$ to the space of test functions is  achieved by \cite{Eckmann73} 
\begin{align} 
 J e^{i {\varphi}(f) } J  = e^{-i {\varphi}(jf) }, \quad \Delta e^{i {\varphi}(f) } \Delta^{-1} = e^{i {\varphi}(\delta f) }. \label{jop}
\end{align} 
Also, it is worth noting that if $f \in \mathcal{M} \implies jf \in \mathcal{M}'$. This property follows from 
\begin{equation} 
s^{\dagger} (jf) = j \delta^{-1/2} jf = \delta f = j (j\delta f) = j (sf) = jf \;. \label{jjf} 
\end{equation} 
An important aspect to be mentioned is that, in the case in which the von Neumann algebra $\mathcal{A'}(\mathcal{M})$ refers to the right wedge ${\cal W}_R$ 
\begin{equation} 
{\cal W}_R = \{\; \{x,y,z,t\} \in{ \mathbb{R}}^4\;; \;x>|t|  \; \} \;, \label{wr}
\end{equation}
the modular operator $\delta$ has a continuous spectrum, given by the positive real line, {\it i.e.}, $\log(\delta) = \mathbb{R}$ \cite{Bisognano75}. To take advantage of this result, in the following, we shall always assume that the region ${\cal O}$ is an open region of ${\cal W}_R$. \\\\We can  now  evaluate  the correlation functions of the Weyl operators:  
\begin{equation}
\langle e^{i \varphi(f_A)} e^{ \pm i\varphi(f_B)}\rangle = \langle e^{i( \varphi(f_A)\pm \varphi(f_B))}\rangle = e^{-\frac{1}{2} ||f_A \pm f_B||^2} \;, \label{Wex}
\end{equation}
where $\{f_A\}$ and $\{f_B\}$ denote two pairs of space-like test functions corresponding, respectively, to  Alice's and Bob's  test functions.  Let us first focus on the pair of Alice's test function $\{f_A\}=(f,f')$. We require that $\{ f_A \}=(f,f')  \in {\cal M(O)}$ where ${\cal O} \in {\cal W}_R$.  Following \cite{SW1,Summers87b,DeFabritiis:2023tkh}, the test function $\{f_A\}=(f,f')$ can be  specified by relying on the spectrum of the operator $\delta$. Picking up the spectral subspace specified by $[\lambda^2-\varepsilon, \lambda^2+\varepsilon ] \subset (0,1)$ and introducing  the normalized vector $\phi$ belonging to this subspace, one writes 
\begin{equation}
f  = \eta  (1+s) \phi \;, \qquad f' = \eta' (1+s) i \phi \;, 
\label{nmf}
\end{equation}
where $(\eta, \eta')$ are free arbitrary parameters, corresponding to the norms of $(f,f')$.  As required by the setup outlined above, equation \eqref{nmf} ensures that 
\begin{equation}
s f = f  \;, \qquad s f'= f' \;.  \label{fafa}
\end{equation}
One notices also that $j\phi$ is orthogonal to $\phi$, {\it i.e.}, $\langle \phi |  j\phi \rangle = 0$. In fact, from 
\begin{align} 
\delta^{-1} (j \phi) =  j (j \delta^{-1} j) \phi = j (\delta \phi), 
\label{orth}
\end{align}
it follows that the modular conjugation $j$ exchanges the spectral subspace $[\lambda^2-\varepsilon, \lambda^2+\varepsilon ]$ into $[1/\lambda^2-\varepsilon,1/ \lambda^2+\varepsilon ]$, ensuring that $\phi$ and $j \phi$ are orthogonal.   Concerning now the pair of Bob's test function $\{f_B\}$, we make use of the modular conjugation operator $j$ and define 
\begin{equation} 
\{f_B\} = (jf, jf')   \;, \label{fb}
\end{equation}
so that 
\begin{equation} 
s^{\dagger}(jf) = jf \;, \qquad s^{\dagger}(jf') = jf'  \;, \label{jffp}
\end{equation}
meaning that, as required by the relativistic causality, $(jf,jf')$ belong to the symplectic  complement $\mathcal{M'}(\mathcal{O})$, located in the left Rindler wedge, namely:$(jf,jf') \in  \mathcal{M'}(\mathcal{O})$. Finally, taking into account that $\phi$ belongs to the spectral subspace $[\lambda^2-\varepsilon, \lambda^2+\varepsilon ] $, it follows that \cite{SW1,Summers87b,DeFabritiis:2023tkh}
\begin{align}
\vert\vert f \vert\vert^2  &= \vert\vert jf  \vert\vert^2 = \eta^2 (1+\lambda^2) \nonumber \\
\langle f \vert jf  \rangle &= 2 \eta^2 \lambda  \nonumber \\
\vert\vert f' \vert\vert^2  &= \vert\vert jf'  \vert\vert^2 = {\eta'}^2 (1+\lambda^2) \nonumber \\
\langle f' \vert  jf' \rangle &= 2 {\eta'}^2 \lambda  \nonumber \\
\langle f \vert jf' \rangle &=  0  \;.  \label{sfl}
\end{align}

\section{Working in ${\cal H} \otimes {\cal H}_{AB}$: construction of Hermitian  dichotomic operators and the Bell-CHSH inequality} 

As mentioned in the Introduction, we shall consider a system whose Hilbert space is ${\cal H} \otimes {\cal H}_{AB}$, where ${\cal H}$ is the Hilbert space of the scalar quantum field and ${\cal H}_{AB}$ is the Hilbert space of a bipartite quantum mechanical system: ${\cal H}_{AB}= {\cal H}_A \otimes {\cal H}_B$ with ${\rm dim}({\cal H}_A) = {\rm dim}({\cal H}_B)= N$ \\\\The first task is that of constructing a von Neumann algebra on ${\cal H} \otimes {\cal H}_{AB}$ with a cyclic and separating state. A concrete realization of the Hilbert space ${\cal H}$ is given by the  Reeh-Schlieder theorem \cite{Haag92,Witten18}. Let $|0\rangle$ be the vacuum state of the field $\varphi$. Thus \cite{Guido:2008jk}: 
\begin{equation} 
{\cal H} = {\rm span}  \{ W_{f_1}.....W_{f_n} |0\rangle \} \label{h}
\end{equation}
where $W_f = e^{i\varphi(f)}$ stands for the Weyl operator and $(f_1,...,f_n)$ are test functions $\in {\cal M}({\cal O})$. Said otherwise, the vacuum state $|0\rangle$ is cyclic and separating for the von Neumann Algebra ${\cal A}({\cal M})$ generated by the Weyl operators $\{ W_f \}$. As for the cyclic and separating state on ${\cal H}_{AB}$, the goal is accomplished by introducing the state 
\begin{equation} 
 |\psi_{AB}\rangle = \sum_{j=1}^N c_j |j\rangle_A \otimes |j \rangle_B \;, \qquad \sum_{j=1}^N |c_j|^2 =1 \;, \qquad c_j \neq 0 \; \forall j \;, \label{pisab} 
\end{equation} 
where $\{| j\rangle_A, j=1,...,N\}$ and $\{| j \rangle_B, j=1,...N \}$ are basis states in ${\cal H}_A$ and ${\cal H}_B$. As shown in details in \cite{Witten18}, the state $|\psi_{AB}\rangle$ is cyclic and separating for the von Neumann algebra 
\begin{equation} 
{\hat {\cal O}} =\{ O_A \otimes {\mathbb{1}}_B \}  \;, \label{vno}
\end{equation} 
where $O_A$ are the operators of ${\cal H}_A$\footnote{The operators $\{O_A \}$ are the $N \times N $ complex matrices.}. \\\\Therefore, the state 
\begin{equation} 
|\psi \rangle = |0 \rangle \otimes |\psi_{AB} \rangle \;, \label{psi} 
\end{equation} 
is cyclic and separating for the von Neumann algebra 
\begin{equation} 
{\hat {\cal A}_O} = {\cal A}({\cal M}) \otimes {\hat {\cal O}}  \;. \label{vnao}
\end{equation}
As outlined in the previous section, the relevance of having a cyclic and separating state for  ${\hat {\cal A}_O}$ relies on the possibility of applying the powerful modular theory of Tomita-Takesaki. 

\subsection{Construction of Bell's Hermitian dichotomic operators on ${\cal H} \otimes {\cal H}_{AB}$. The Bell-CHSH inequality}

We focus now on the construction of the Bell Hermitian dichotomic operators $(A,A'.B.B')$ needed for the formulation of the Bell-CHSH inequality on ${\cal H} \otimes {\cal H}_{AB}$. Without loss og generality, we may take $N$ even, {\it i.e} $N=2M$, M integer. To introduce Alice's  operators $(A,A')$ we define 
\begin{equation} 
A | i \rangle_A = e^{i\varphi(f)}\;|i+1\rangle_A  \;, \qquad  A | i+1 \rangle_A = e^{-i\varphi(f)}\; |i\rangle_A  \;, \label{aop}
\end{equation}
\begin{equation} 
A' | i \rangle_A = e^{i\varphi(f')}\;|i+1\rangle_A  \;, \qquad  A' | i+1 \rangle_A = e^{-i\varphi(f')}\; |i\rangle_A  \;, \label{apop}
\end{equation}
where $(f,f')$ denote Alices's test functions, as introduced in the previous section. Making use of the modular conjugation operator $j$, eq.\eqref{jd}, for Bob's operators $(B,B')$, one has 
\begin{equation} 
B | i \rangle_B = e^{i\varphi(jf)}\;|i+1\rangle_B  \;, \qquad  B | i+1 \rangle_B = e^{-i\varphi(jf)}\; |i\rangle_B  \;, \label{bop}
\end{equation}
and
\begin{equation} 
B' | i \rangle_B = e^{i\varphi(jf')}\;|i+1\rangle_B  \;, \qquad  B' | i+1 \rangle_B = e^{-i\varphi(jf')}\; |i\rangle_B  \;, \label{bpop}
\end{equation}
where $(jf,jf')$ are Bob's test functions. \\\\One easily checks that the operators $(A,A')$, $(B,B')$ fulfill all needed requirements, that is:
\begin{eqnarray} 
A^{\dagger} & = & A \;, \qquad A^2 =1 \;, \qquad A'^{\dagger} = A' \;, \qquad A'^2=1 \nonumber \\
B^{\dagger} & = & B \;, \qquad B^2 =1 \;, \qquad B'^{\dagger} = B' \;, \qquad B'^2=1 \;, \label{abb}
\end{eqnarray} 
with 
\begin{equation}
 [  A ,B]  = [A,B']= [A',B]= [A,B'] = 0 \;. \label{abb1}
\end{equation} 
The quadruple $(A,A',B,B')$ is thus an eligible set of operators for the Bell-CHSH inequality in ${\cal H} \otimes {\cal H}_{AB}$. 

\subsection{The Bell-CHSH inequality in the vacuum state of the quantum field}

In order to establish the Bell-CHSH inequality for the quantum field, we start by considering the Bell-CHSH Hermitian operator ${\cal C}$ in ${\cal H} \otimes {\cal H}_{AB}$, namely 
\begin{equation} 
{\cal C} = (A + A') \otimes B + (A-A') \otimes B'  \;, \label{cc}
\end{equation}
and consider its correlation function in the state $|\psi\rangle $: 
\begin{equation} 
\langle {\cal C} \rangle_{\psi} = \langle \psi | \; (A + A') \otimes B + (A-A') \otimes B' \; | \psi \rangle \;. \label{cc1}
\end{equation}
Since $(A,A',B,B')$ are Hermitian and dichotomic, Tsirelson's argument \cite{tsi1} applies, implying that 
\begin{equation} 
|| {\cal C} || \le 2 \sqrt{2} \;, \label{ts}
\end{equation} 
where $||{\cal C}|| $ is the operator norm of ${\cal C}$. \\\\Let us now consider the operator ${\cal C}_0$ acting on the Hilbert space of the quantum field ${\cal H}$, obtained by working out the expectation value of ${\cal C}$ in the Hilbert space  ${\cal H}_{AB}$ of the quantum mechanical system, {\it i.e.} 
\begin{equation} 
{\cal C}_0 = \langle \psi_{AB} \; | {\cal C} \;| \psi_{AB} \rangle \;. \label{cnot}
\end{equation}
From equation \eqref{ts}, it turns out that 
\begin{equation} 
| \langle 0 | \; {\cal C}_0 \; |  0 \rangle  | \le 2 \sqrt{2} \;. \label{tss}
\end{equation}
Moreover, an elementary calculations shows that 
\begin{equation} 
\langle \psi_{AB} \;| A\otimes B \;| \psi_{AB} \rangle = \cos(\varphi(f) + \varphi(jf)) \;. \label{cs1}
\end{equation} 
Therefore, taking into account that the vacuum expectation value of an odd number of fields vanishes, for the Bell-CHSH of the quantum field in the vacuum state, we get 
\begin{equation} 
\langle  {\cal C}_0\rangle = \langle 0|\; (W_f+ W_{f'})W_{jf} + (W_f -W_{f'})W_{jf'} \;|0\rangle =\langle 0|\; ( e^{i\varphi(f)} + e^{i\varphi(f')} ) 
e^{i \varphi(jf) }  +  (e^{i\varphi(f)} - e^{i\varphi(f')} )  e^{i \varphi(jf) }\; |0\rangle  \label{bco}
\end{equation}
One speaks of a violation of the Bell-CHSH inequality in the vacuum state of a scalar quantum field whenever 
\begin{equation} 
2 < | \langle  {\cal C}_0\rangle | \le 2 \sqrt{2} \;. \label{vb}
\end{equation} 
Expression \eqref{bco}  is precisely the Bell-CHSH inequality derived in \cite{Guimaraes:2024alk}, relying on the use of unitary operators. One sees thus that employing the Hilbert space ${\cal H} \otimes {\cal H}_{AB}$, the same inequality of \cite{Guimaraes:2024alk} follows in a simple and elegant way by introducing the Hermitian dichotomic field dependent operators, eqs.\eqref{aop}-\eqref{bpop}. These  operators  can be regarded as the natural generalization to Quantum Field Theory of the usual Bell's spin operators of Quantum Mechanics. \\\\Using the inner products of eqs.\eqref{sfl}, for $\langle {\cal C}_0\rangle$ one gets 
\begin{equation} 
\langle {\cal C}_0\rangle = e^{-\eta^2(1+\lambda)^2} + 2 e^{-\frac{1}{2}(\eta^2+ \eta'^2)(1+\lambda^2)} - e^{-\eta'^2(1+\lambda)^2}  \;. \label{vv}
\end{equation}
In order to have a concrete idea of the size of the violation achieved by eq.\eqref{vv}, one might employ the following choice 
\begin{equation} 
\eta= 0.01 \;, \qquad \eta'= 0.564058 \;, \qquad \lambda = 0.495456 \;, \label{values}
\end{equation} 
resulting in 
\begin{equation} 
\langle {\cal C}_0\rangle = 2.14931 \;. \label{vt}
\end{equation}
Let us end this section by mentioning that the inequality \eqref{vb} can be derived also in the case in which the Hilbert space ${\cal H}_{AB}$ has infinite dimension. Let us illustrate this point by considering two oscillators. For the basis states we have 
\begin{equation} 
\{ |n_A\rangle = \frac{(a^{\dagger})^{n}}{\sqrt{n !}} |0\rangle_A, \;n=0,1...., \}\;, \qquad \{ |m_B\rangle = \frac{(b^{\dagger})^{m}}{\sqrt{m !}} |0\rangle_B, \;m=0,1...., \}\;, \label{basis}
\end{equation}
with $(a,b)$ fulfilling the commutation relations 
\begin{eqnarray} 
[a, a^\dagger] & = & 1\;, \qquad [a,a]=[a^\dagger, a^\dagger] = 0 \;, \nonumber \\
\left[ b, b^\dagger  \right] & = & 1\;, \qquad [b,b]=[b^\dagger, b^\dagger] = 0 \;, \nonumber \\
\left[ a,b \right] & = & [a, b^\dagger] =0 \;. \label{ccr}
\end{eqnarray}
In this case, the state $|\psi_{AB}\rangle$ is taken to be the squeezed state 
\begin{equation} 
| \psi_{AB} \rangle = \left(1 -\delta^2\right)^{1/2} \sum_{n=0}^{\infty} \delta^{n} |n_A\rangle \otimes |n_B \rangle \;, \qquad \langle \psi_{AB} | \psi_{AB} \rangle = 1 \;. \label{squ} 
\end{equation}
where $\delta \in [0,1]$ is the squeezing parameter. For the operators $(A,A',B,B')$ one writes 
\begin{equation} 
A | 2n_A \rangle = e^{i\varphi(f)}\;|2n_A+1\rangle  \;, \qquad  A | 2n_A+1 \rangle  = e^{-i\varphi(f)}\; |2n_A\rangle  \;, \label{aop1}
\end{equation}
\begin{equation} 
A' |2n_A \rangle = e^{i\varphi(f')}\;|2n_A+1\rangle  \;, \qquad  A' | 2n_A+1 \rangle = e^{-i\varphi(f')}\; |2n_A\rangle \;, \label{apop1}
\end{equation}
and
\begin{equation} 
B | 2n_B\rangle = e^{i\varphi(jf)}\;|2n_B+1\rangle  \;, \qquad  B | 2n_B+1 \rangle = e^{-i\varphi(jf)}\; |2n_B\rangle  \;, \label{bop1}
\end{equation}
\begin{equation} 
B' | 2n_B \rangle = e^{i\varphi(jf')}\;|2n_B+1\rangle  \;, \qquad  B' | 2n_B+1 \rangle = e^{-i\varphi(jf')}\; |2n_B\rangle  \;. \label{bpop1}
\end{equation}
Evaluating $\langle \psi_{AB}\; | A \otimes B\;| \psi_{AB} \rangle $, one gets   
\begin{equation} 
\langle \psi_{AB}\; | A \otimes B\;| \psi_{AB} \rangle = \frac{2 \delta}{1+\delta^2} \cos(\varphi(f)) + \varphi(jf))  \;, \label{evs}
\end{equation} 
and taking the limit $\delta \rightarrow 1$, the Bell-CHSH inequality \eqref{bco} follows. 

\section{A Jaynes-Cummings type Hamiltonian. Evaluation of the perturbative corrections to the Bell-CHSH inequality}\label{JC}

In this section we employ the previous construction to study the quantum corrections to the Bell-CHSH inequality via a Jaynes-Cummings type Hamiltonian describing the interaction between the scalar field and a pair of qbits. More precisely, we shall start with the unperturbed Hamiltonian 
\begin{equation} 
H_0 = H_s + \int d \mu_p \omega_p a^{\dagger}_p a_p  \;, \label{hmo}
\end{equation}
where $H_s$ stands for the Heisembenrg Hamiltonian 
\begin{equation} 
H_s = J \left( \sigma^A_x \otimes \sigma^B_x + \sigma^A_y \otimes \sigma^B_y+ \sigma^A_z \otimes \sigma^B_z \right) \;, \label{hss}
\end{equation}
for two spins $1/2$. \\\\The ground state of $H_s$ is the Bell singlet state 
\begin{equation} 
|\psi_s\rangle = \frac{|+\rangle_A \otimes |-\rangle_B - | -\rangle_A\otimes |+\rangle_B}{\sqrt{2}} \;, \label{sg}
\end{equation} 
with 
\begin{equation} 
H_s |\psi_s\rangle = - 3J |\psi_s\rangle \;.\label  {j3} 
\end{equation}
The second term in eq.\eqref{hmo} is the Hamiltonian of the free field: 
\begin{equation} 
d\mu_p = \frac{d^3p}{(2\pi)^3} \frac{1}{2 \omega_{p}} \;, \qquad [a_p, a^{\dagger}_q ] = (2 \pi)^3 2 \omega_p \delta^3(p-q) \;. \label{qtf}
\end{equation}
The ground state of $H_0$ is 
\begin{equation} 
|\psi_g\rangle = |\psi_s \rangle \otimes  |0 \rangle 
\end{equation}
We now perturb $H_0$ by introducing the following Jaynes-Cummings type interaction Hamiltonian 
\begin{equation}
H_I = \Omega_A \left( \sigma^A_+ a_{h_A} + \sigma^A_{-} a^{\dagger}_{h_A} \right) + \Omega_B \left( \sigma^B_+ a_{h_B} + \sigma^B_{-} a^{\dagger}_{h_B} \right) \;, \label{hint}
\end{equation}
where $(\Omega_A, \Omega_B)$ are coupling constants and 
\begin{equation} 
\sigma_{+} =\frac{\sigma_x + i \sigma_y}{2} \;, \qquad \sigma_{-} =\frac{\sigma_x + i \sigma_y}{2}
\;. \label{ss}
\end{equation}
Also
\begin{equation} 
a_{h_A} = \int d\mu_p \; h_A(\omega_p, {\vec p}) a_p \;, \qquad  a_{h_B} = \int d\mu_p \; h_B(\omega_p, {\vec p}) a_p \;, \label{aaff}
\end{equation}
are the smeared annihilation operators corresponding to the test functions $(h_A,h_B)$, whose respective supports are space-like. Notice that, due to the form of $H_I$, only the vacuum state $|0\rangle$ and the 1-particle states $\{ |p \rangle = a^{\dagger}_p |0\rangle \}$ do matter.  \\\\In order to apply perturbation theory and evaluate the second order corrections to the wave function $|\psi_g\rangle$, we shall employ the Bell orthonormal basis for the qbits, namely $\{|\psi_s\rangle, |\psi_j\rangle, j=1,2,3\}$, with $|\psi_s\rangle$ given in eq.\eqref{sg} and: 
\begin{eqnarray} 
|\psi_1\rangle & = & \frac{|+\rangle_A \otimes |-\rangle_B + | -\rangle_A\otimes |+\rangle_B}{\sqrt{2}} \nonumber \\
|\psi_2\rangle & = & \frac{|+\rangle_A \otimes |+\rangle_B + | -\rangle_A\otimes |=\rangle_B}{\sqrt{2}} \nonumber \\
|\psi_3\rangle & = & \frac{|+\rangle_A \otimes |+\rangle_B - | -\rangle_A\otimes |=\rangle_B}{\sqrt{2}} \;. \label{Bbs}
\end{eqnarray} 
A  straightforward use of perturbation theory gives the second-order corrected wave function as 
\begin{eqnarray} 
|\psi^{(2)}_g \rangle  & = & \left( 1 - \int d\mu_p \frac{|\Omega_A h_A(p) - \Omega_B h_B(p)|^2}{4(4J + \omega_p)^2} \right) |\psi_s\rangle \otimes |0\rangle \nonumber \\
&-& \int d \mu_p \left( \frac{\left( \Omega_A h_A(p) -\Omega_B h_B(p)\right)}{2 (4J+\omega_p)}\right) \left( |\psi_2\rangle -|\psi_3\rangle \right) \otimes |p \rangle \nonumber \\
&+&  \int d \mu_p \left( \frac{\left( \Omega_A^2 h_A^2(p) -\Omega_B^2 h^2_B(p)\right)}{8J (4J+\omega_p)}\right)  |\psi_1\rangle  \otimes |0 \rangle \;. \label{sec}
\end{eqnarray}
Setting 
\begin{eqnarray} 
A |+\rangle_A & = & e^{i\varphi(f)} |-\rangle_A \;, \qquad  A |-\rangle_A = e^{-i\varphi(f)} |+\rangle_A \nonumber \\
B|+\rangle_B & = & e^{i\varphi(jf)} |-\rangle_B \;, \qquad  B |-\rangle_A = e^{-i\varphi(jf)} |-\rangle_B  \;, \label{pm}
\end{eqnarray} 
and similar equations for $(A',B')$, for the second order quantum corrections to the Bell-CHSH inequality, we get 
\begin{equation} 
\langle \psi^{(2)}_g \;| {\cal C} \; |\psi^{(2)}_g \rangle = \left(1 -\delta^2 \right)  \langle {\cal C}_0 \rangle \;, \label{uu}
\end{equation}
where $\langle {\cal C}_0 \rangle$ as in eq.\eqref{vv} and $\delta^2$ given by 
\begin{equation} 
\delta^2 = \int d\mu_p \frac{|\Omega_A h_A(p) - \Omega_B h_B(p)|^2}{2(4J + \omega_p)^2} \;. \label{dd}
\end{equation} 
Equation \eqref{dd} displays a nice feature of the Bell-CHSH inequality. One notices that the first order corrections, {\it i.e} terms linear in $\Omega_A$ and $\Omega_B$, vanish and that the second order term has a definite sign, namely $\delta^2>0$. As such, expression \eqref{uu} is perfectly compatible with Tsirelson's bound Suppose in fact that suitable spacetime localization regions for Alice and Bob have been found in such a way that $\langle {\cal C}_0 \rangle$ achieves its maximum, {\it i.e.} $\langle {\cal C}_0 \rangle = 2 \sqrt{2}$. Since Tsirelson's bound is the maximum allowed value, perturbative corrections can only have the effect of producing a decrease of the violation, as shown in fact by equation \eqref{uu}.

\section{ Further considerations and examples }

In order to provide a better understanding of the setup presented in the previous sections, we add  a few considerations by making use of Quantum Mechanical examples. \\\\A feature which we highlight is that the wave function, eq.\eqref{psi}, has the form 
\begin{equation}
|\psi\rangle = |\psi_1\rangle \otimes |\psi_2\rangle \;, \label{tp}
\end{equation}
where $|\psi_1\rangle$ and $|\psi_2\rangle$ are highly entangled states. It is worth reminding here that the fact that the vacuum state $|0\rangle$ of a scalar quantum field $\varphi$ is entangled is well established \cite{SW1,Summers87b,Summers:1987ze}. Although the wave function $|\psi\rangle$ 
is a tensor product of two states, one can argue that the entanglement properties of  $|\psi_1\rangle$ and $|\psi_2\rangle$ are lifted to the whole state. This seems to be a general property of the states of the form given in eq.\eqref{tp}, as it can be illustrated with the help of the following simple example, in which both Alice and Bob hold a composite system made up by a spin $1$ and a spin $1/2$ particles. \\\\In this case, a basis for Alice's Hilbert space is given by the six states $(|1\rangle_A |+\rangle_A, |-1\rangle_A |+\rangle_A, |0\rangle_A |+\rangle_A, |1\rangle_A |-\rangle_A, |-1\rangle_A |-\rangle_A, |0\rangle_A |-\rangle_A)$. The same for Bob's basis. For the Hermitian dichotomic Bell operators $(A,A'.B.B')$, we take 
\begin{eqnarray} 
A |1\rangle_A |+\rangle_A & = & e^{i(\alpha_1 + \alpha_2)} |-1\rangle_A |-\rangle_A \;, \qquad A |-1\rangle_A |+\rangle_A  =  e^{-i(\alpha_1 - \alpha_2)} | 1\rangle_A |-\rangle_A\;,  \nonumber \\
A |0\rangle_A |+\rangle_A & = & e^{i \alpha_2} |0\rangle_A |-\rangle_A \;, \qquad \;\;\;\; \;\;\;\;\;\;\;\;A |1\rangle_A |-\rangle_A  =  e^{i(\alpha_1 - \alpha_2)} |-1\rangle_A |+\rangle_A  \;, \nonumber \\
A |-1\rangle_A |-\rangle_A & = & e^{-i(\alpha_1 + \alpha_2)} |1\rangle_A |+\rangle_A \;, \qquad \;\;A |0\rangle_A |-\rangle_A  =  e^{-i \alpha_2 } |0\rangle_A |+\rangle_A \;. \label{a112}
\end{eqnarray} 
and 
\begin{eqnarray} 
B |1\rangle_B |+\rangle_B & = & e^{i(\beta_1 + \beta_2)} |-1\rangle_B |-\rangle_B \;, \qquad B |-1\rangle_B |+\rangle_B  =  e^{-i(\beta_1 - \beta_2)} | 1\rangle_B |-\rangle_B\;,  \nonumber \\
B |0\rangle_B |+\rangle_B & = & e^{i \beta_2} |0\rangle_B |-\rangle_B \;, \qquad \;\;\;\; \;\;\;\;\;\;\;\;B |1\rangle_B |-\rangle_B  =  e^{i(\beta_1 - \beta_2)} |-1\rangle_B |+\rangle_B  \;, \nonumber \\
B |-1\rangle_B |-\rangle_B & = & e^{-i(\beta_1 + \beta_2)} |1\rangle_B |+\rangle_B \;, \qquad \;\;B |0\rangle_B |-\rangle_B  =  e^{-i \beta_2 } |0\rangle_B |+\rangle_B \;.,\label{b112}
\end{eqnarray} 
where $(\alpha_1,\alpha_2)$ and $(\beta_1, \beta_2)$ are arbitrary parameters. The expressions for $(A', B')$ are obtained from those of $(A,B)$ upon replacing $(\alpha_1,\alpha_2) \rightarrow (\alpha_1',\alpha_2')$ and $(\beta_1, \beta_2)\rightarrow (\beta_1', \beta_2')$. It turns out that these operators obey the whole set of conditions \eqref{abb},\eqref{abb1}.\\\\We can now consider two cases:
\begin{itemize} 
\item in the first case we have $ |\psi\rangle = |\psi_1\rangle \otimes |\psi_2\rangle $, where 
\begin{eqnarray} 
|\psi_1\rangle & = & \frac{1}{\sqrt{3}} \left( |1\rangle_A |-1\rangle_B - |0\rangle_A |0\rangle_B + |-1\rangle_A |1\rangle_B \right)  \;, \nonumber \\[3mm]
|\psi_2\rangle & = & \frac{1}{\sqrt{2}} \left( |+\rangle_A |-\rangle_B - |-\rangle_A |+ \rangle_B \right)  \;, \label{s112}
\end{eqnarray}
are the entangled singlet states for spin $1$ and spin $1/2$, respectively. For the matrix element $\langle \psi |\; A\otimes B \; |\psi \rangle$  one gets 
\begin{equation} 
\langle \psi |\; A\otimes B \; |\psi \rangle = - \frac{1}{3} ( 1 + 2 \cos(\alpha_1-\beta_1)) \cos(\alpha_2 - \beta_2)  \;,.\label{sab}
\end{equation}
It follows that the Bell-CHSH correlator turns out to be 
\begin{eqnarray} 
\Big| \langle \psi |\; {\cal C} \; |\psi \rangle \Big| & =& \frac{1}{3}\Big|  ( 1 + 2 \cos(\alpha_1-\beta_1)) \cos(\alpha_2 - \beta_2) +  ( 1 + 2 \cos(\alpha_1'-\beta_1)) \cos(\alpha_2 - \beta_2) \nonumber  \\
& \;\;\;& + ( 1 + 2 \cos(\alpha_1-\beta_1')) \cos(\alpha_2 - \beta_2) -  ( 1 + 2 \cos(\alpha_1'-\beta_1')) \cos(\alpha_2' - \beta_2') \Big| \;. \label{Cs12}
\end{eqnarray} 
Setting $(\alpha_2=0, \alpha_2'= \frac{\pi}{2}, \beta_2 = \frac{\pi}{4}, \beta_2'=-\frac{\pi}{4})$ and  $(\alpha_1=0, \alpha_1'= \frac{\pi}{2}, \beta_1 = \frac{\pi}{4}, \beta_1'=\frac{\pi}{4})$, one has 
\begin{equation} 
\Big| \langle \psi |\; {\cal C} \; |\psi \rangle \Big| = \frac{2 \sqrt{2} (1 + \sqrt{2})}{3} \approx 2.27  \;, \label{CCC}
\end{equation}
resulting in a violation of the Bell-CHSH inequality. 
\item in the second case, one of the two states $(|\psi_1\rangle, |\psi_2\rangle)$ is a product state, for instance:
\begin{equation} 
|\psi_1\rangle  =   |1\rangle_A |-1\rangle_B \;. \label{pdd}
\end{equation} 
Unlike the first case, the state $|\psi_1\rangle$ is not entangled. An elementary calculation shows that in this second case  no violation  takes place. 
\end{itemize}
We see therefore that, when both $(|\psi_1\rangle, |\psi_2\rangle)$ are entangled, one ends up with a violation of the Bell-CHSH inequality. The entanglement of both states $(|\psi_1\rangle, |\psi_2\rangle)$ is needed to have violation. If one of the two states $(|\psi_1\rangle, \psi_2\rangle)$ is not entangled, the violation is lost. \\\\Let us conclude this section by remarking that the form of the Hermitian dichotomic Bell operators given in eqs.\eqref{aop}-\eqref{bpop} cannot be thought as the product of two factors, the first one acting on the quantum field and the second one on th spin $1/2$ states. The operators of  eqs.\eqref{aop}-\eqref{bpop} feel the full structure of the state $|0\rangle \otimes |\psi_{AB}\rangle$. As such, these operators are useful to probe the correlations existing in the system as a whole.

\section{Conclusion} 
	
In this work, the Bell-CHSH inequality has been studied by employing the Hilbert space 	${\cal H} \otimes {\cal H}_{AB}$, where ${\cal H}$ and ${\cal H}_{AB}$ stand for the Hilbert space of a scalar quantum field and of a generic bipartite quantum mechanical system. As outlined in the text, there are several advantages in working with ${\cal H} \otimes {\cal H}_{AB}$. \\\\In particular, the construction of Hermitian, field dependent, dichotomic Bell's operators becomes simple and elegant, as expressed by eqs.\eqref{aop}-\eqref{bpop}. Moreover, the formulation of the Bell-CHSH inequality in the vacuum state of the scalar field in terms of the unitary Weyl operators emerges in a natural and clever way, eq.\eqref{bco}. \\\\The use of the Hilbert space ${\cal H} \otimes {\cal H}_{AB}$ enables  the investigation of the issue of the existence of quantum corrections to the Bell-CHSH inequality, via the use of Jaynes-Cummings type Hamiltonians. The output of the second order calculation performed in Sec.\eqref{JC}  shows in an explicit way that the quantum corrections are, as expected,  compatible with Tsirelson's bound. 

\section*{Acknowledgments}
	The authors would like to thank the Brazilian agencies CNPq and CAPES, for financial support.  S.P.~Sorella, I.~Roditi, and M.S.~Guimaraes are CNPq researchers under contracts 301030/2019-7, 311876/2021-8, and 310049/2020-2, respectively.

\end{document}